\documentclass[aip,jcp,11pt,reprint,floatfix,superscriptaddress,urlname]{revtex4-1}
\usepackage[utf8]{inputenc}
\usepackage{amsmath}
\usepackage{color}
\usepackage{graphicx}
\usepackage{comment}
\usepackage{caption}
\usepackage{subcaption}
\usepackage[normalem]{ulem}
\usepackage{algorithm}
\usepackage{algpseudocode}

\usepackage{amsmath,amssymb}
\usepackage{graphicx}
\usepackage{booktabs}

\usepackage{xcolor}

\newcommand{\ket}[1]{{\big| #1 \big>}}

\DeclareMathAlphabet      {\mathbfit}{OML}{cmm}{b}{it}

\def\mi{\mathrm{i}}

\begin{document}

\title{Exact relationships between the GW approximation and equation-of-motion coupled-cluster theories through the quasi-boson formalism}

\author{Johannes T\"olle}
\email{jtolle@caltech.edu}
\author{Garnet Kin-Lic Chan}
\email{gkc1000@gmail.com}
\noaffiliation
\affiliation{Division of Chemistry and Chemical Engineering, California Institute of Technology, Pasadena, California 91125, USA}

\begin{abstract}
We describe the relationship between the GW approximation and various equation-of-motion (EOM) coupled-cluster (CC) theories. We demonstrate the exact equivalence of the G$_0$W$_0$ approximation and the propagator theory for an electron-boson problem in a particular excitation basis. From there, we establish equivalence within the quasi-boson picture to the IP+EA-EOM unitary coupled-cluster propagator. We analyze the incomplete description of screening provided by the standard similarity-transformed IP+EA-EOM-CC and the recently introduced G$_0$W$_0$ Tamm-Dancoff approximation. We further consider the approximate decoupling of IP and EA sectors in EOM-CC treatments and devise the analogous particle-hole decoupling approach for the G$_0$W$_0$ approximation. Finally, we numerically demonstrate the exact relationships and magnitude of the  approximations in calculations of a set of molecular ionization potentials and electron affinities. 
\end{abstract}
\maketitle

\section{Introduction}
The GW approximation, an approximation to the electron self-energy, is widely used to compute quasiparticle electron energies beyond the level of Kohn-Sham density functional theory in both molecules and materials. Although originally defined as a self-consistent self-energy approximation where the Green's function and self-energy are iterated to consistency, it is commonly used in its one-shot formulation, denoted G$_0$W$_0$. Despite many notable successes, the GW approximation has proven hard to improve upon while retaining its favourable computational features \cite{lewis2019vertex,bruneval2021gw}, and furthermore removing the starting point dependence of G$_0$W$_0$ remains a practical challenge. For an overview of current developments we refer to the review of Ref.~\citenum{golze2019gw}.
\\
In parallel, time-independent wavefunction approaches for the determination of quasiparticle energies have seen much development in molecular quantum chemistry~\cite{rittby1988open,nooijen1993coupled,nooijen1995equation,schirmer1998non,chatterjee2019second,banerjee2019third,dempwolff2022vertical}, and have more recently been implemented for materials~\cite{katagiri2005equation,mcclain2017,furukawa2018band,zhang2019coupled,gallo2021periodic,lange2021improving,banerjee2022non}. 
One such widely used set of methods is the 
equation-of-motion (EOM) coupled-cluster (CC) theory~\cite{monkhorst1977calculation,koch1990excitation,stanton1993equation}, whose variant for quasiparticle energies, i.e.~the ionization potentials (IP) and electron affinities (EA), is termed IP/EA-EOM-CC.
These methods start from the ground-state CC theory, and the EOM approximation can be related to the linear-response of the ground-state ansatz. 
The IP/EA-EOM-CC framework offers a simple hierarchy of approximations that become increasingly exact.
Unfortunately, even at the lowest commonly employed level of approximation, namely the single and doubles level (IP/EA-EOM-CCSD), the computational cost typically exceeds that of standard numerical implementations of the GW approximation\cite{hybertsen1986electron,godby1988self,rieger1999gw,golze2018core,zhu2021all}. 
\\
The popularity of both the GW and EOM-CC approaches has led to recent interest in examining the connections between the two. For example, $\mathrm{G}_0\mathrm{W}_0$ is based on a random phase approximation (RPA) description of screening which may also be used to describe ground-state correlations, and the connection between the so-called ``direct ring'' version of coupled-cluster theory and the RPA ground-state has been known for many years~\cite{freeman1977coupled, gruneis2009making, scuseria2008ground,jansen2010equivalence,scuseria2013particle,peng2013equivalence}. 
Additionally, the connection between RPA (neutral) excitation energies and EOM-CC has been established \cite{emrich1981extension,berkelbach2018communication,rishi2020route}.
Refs.~\citenum{lange2018relation, quintero2022connections} have analyzed and introduced a variety of useful analogies and resemblances between GW and different IP/EA-EOM-CC approximations. However, these prior works have stopped short of establishing any exact mappings between GW and EOM approaches, complicating the understanding of the relationship between the theories. 
\\
Here, we provide an analysis of the GW and IP/EA-EOM-CC theories that goes beyond previous results in the following aspects:
\begin{enumerate}
    \item We demonstrate an exact equivalence between $\mathrm{G}_0\mathrm{W}_0$ and the EOM theory for an electron-boson problem in a particular excitation basis. In particular we write down the precise time-independent Hamiltonian supermatrix whose resolvent yields the $\mathrm{G}_0\mathrm{W}_0$ Green's function. Although this expression can be deduced following similar arguments to those used in Refs.~\citenum{bintrim2021full,quintero2022connections}, it goes beyond the explicit analysis in such work which 
    establishes a similar equivalence  only within the Tamm-Dancoff approximation (TDA) to screening \cite{bintrim2021full,monino2022unphysical}. The frequency-independent matrix construction we demonstrate is closely related to techniques used to construct the bath in extended dynamical mean-field theory \cite{si1996kosterlitz,sun2002extended}, and is an expression of the well-known description of GW as an electronic polaron theory \cite{hedin1999correlation}.
    \item We establish that, within the quasi-boson formalism, the EOM propagator starting from the ring unitary CC doubles (uCC) ground-state  is exactly equivalent to that of $\mathrm{G}_0\mathrm{W}_0$. 
    In contrast, the similarity transform of conventional CC theory does not lead to such an equivalence. This is due to missing some of the non-TDA screening terms, as discussed in Ref.~\citenum{lange2018relation}. We note that general relationships between unitary transformations and propagator theories have been identified in early work, such as in Refs.~\citenum{prasad1985some,mukherjee1989effective,datta1993consistent}. 
    \item The standard EOM-CC formulation neglects the coupling of the IP and EA sectors in the EOM propagator. This is possible due to the use of a partial decoupling transformation that is equivalent to the ground-state CC amplitude equations, as discussed in Ref.~\citenum{nooijen1992coupled}. We identify the analogous decoupling transformation in the quasi-boson formulation and the form of the remaining IP/EA couplings in both the standard (similarity-transform) and unitary CC languages. 
    \item We numerically demonstrate the equivalences and establish the magnitudes of the approximations described above, including the equivalence of the $\mathrm{G}_0\mathrm{W}_0$ and the uCC quasi-boson propagator, the size of the non-TDA contributions to the screening, including those omitted from the standard CC quasi-boson propagator, and the contributions from the residual IP/EA coupling omitted in the standard formulation of EOM-CC theories. 
    \end{enumerate}

\section{Background on the GW approximation}

\subsection{Green's functions and the self-energy}

The central quantity in this work is the interacting one-electron Green's
function. Assuming an orbital basis, the Green's function is represented by a frequency dependent matrix $\mathbf{G}(\omega)$ \cite{fetter_walecka}. The non-interacting one-particle Green's function $\mathbf{G}_0(\omega)$ is defined from an (arbitrary) one-particle operator $\hat{h}$, 
\begin{align}
    \mathbf{G}_0(\omega) = [\omega \mathbf{1} - \mathbf{h}]^{-1}
\end{align}
and the Dyson equation~\cite{hedin1965new} formally relates the non-interacting and interacting Green's functions via the self-energy $\pmb{\Sigma}(\omega)$
\begin{align}
	\mathbf{G}(\omega) = \mathbf{G}_0(\omega) + \mathbf{G}_0(\omega)\pmb{\Sigma}(\omega)\mathbf{G}(\omega)
	\label{eq:G}
\end{align}
or, in terms of $\mathbf{h}$, 
\begin{align}
    \mathbf{G} = [\omega \mathbf{1} - \mathbf{h} - \mathbf{\Sigma}]^{-1}.\label{eq:dysonh}
\end{align}
The self-energy is thus the central quantity to approximate in a many-body calculation.

The interacting Green's function determines the single-particle excitation energies and total electronic energy of the system. The excitations correspond to the poles of $\mathbf{G}(\omega)$ and can be obtained by solving
\begin{align}
	\left[\mathbf{h} + \pmb{\Sigma}(\mathbf{E}) \right] \mathbf{Z} = \mathbf{Z} \mathbf{E}.
	\label{eq:QPDetermination}
\end{align}
The self-energy is commonly divided into a frequency-independent and frequency-dependent (correlation) piece
\begin{align}
    \mathbf{\Sigma}(\omega) = \mathbf{\Sigma}^\mathrm{static} + \mathbf{\Sigma}^c(\omega).
\end{align}
To define $\mathbf{\Sigma}^\mathrm{static}$, we first specify the
 electronic Hamiltonian
\begin{align}
    \hat{H} &= \hat{t} + \hat{V}\notag \\
    &=\sum_{pq} t_{pq} \hat{a}^\dag_p \hat{a}_q + \frac{1}{2} \sum_{prqs} V_{prqs} \hat{a}^\dag_p \hat{a}^\dag_q \hat{a}_s \hat{a}_r \notag \\
    &=\sum_{pq} h_{pq} \hat{a}^\dag_p \hat{a}_q - \sum_{pq} v_{pq} \hat{a}^\dag_p \hat{a}_q + \frac{1}{2} \sum_{prqs} V_{prqs} \hat{a}^\dag_p \hat{a}^\dag_q \hat{a}_s \hat{a}_r\label{eq:hamiltonian}
\end{align}
where $\hat{a}^\dag, \hat{a}$ are electron creation/annihilation operators with respect to the bare electron vacuum and $p,q,r,s$ are electron orbital labels, and we assume all matrix elements are real. In the second line, the operator $\hat{h}$ is the one appearing in the definition of $\mathbf{G}_0$, and is related to
the bare one-electron Hamiltonian by an effective one-electron operator, i.e. $\hat{h} = \hat{t} + \hat{v}$. (For example, if $\hat{h}$ is chosen to be the Kohn-Sham Hamiltonian, then $\hat{v} = \hat{J} + \hat{v}_{xc}$, i.e. the Kohn-Sham effective potential which is the sum of the Coulomb and exchange-correlation potentials evaluated at the Kohn-Sham density). Then
\begin{align}
    \mathbf{\Sigma}^\mathrm{static} = \mathbf{J} + \mathbf{K} - \mathbf{v} 
\end{align}
where $\mathbf{J}$, $\mathbf{K}$ are the Hartree and exchange matrices associated with the density matrix of the current approximation to the Green's function, $\mathbf{D} = -\frac{1}{\pi} \int_{-\infty}^\mu d\omega \ \mathrm{Im} \ \mathbf{G}(\omega)$ ($\mu$ is the chemical potential to satisfy the trace condition $\mathrm{Tr} \ \mathbf{D} = N$, the number of electrons), i.e.
\begin{align}
    J_{qs} = \sum_{pr} V_{prqs} D_{pr} \notag\\
    K_{rq} = \sum_{ps} V_{prqs} D_{ps}.
\end{align}
Note that if $\hat{h}$ is chosen as the Fock operator $\hat{f} = \hat{t} + \hat{J} + \hat{K}$ then $\pmb{\Sigma}^\mathrm{static}=0$. In this work, we will always assume this choice of $\hat{h}$ for simplicity. 
It remains then to define the correlation self-energy $\pmb{\Sigma}^c(\omega)$ which is the principle content of the GW approximation.
\\
As seen from the above, $\pmb{\Sigma}$ depends on the Green's function i.e. $\pmb{\Sigma} [\mathbf{G}]$. When computing it self-consistently, in the first iteration, we have $\pmb{\Sigma}[\mathbf{G}_0]$. A fully self-consistent self-energy leads to important formal properties of the resulting Green's function, such as conservation laws~\cite{baym1961conservation}. However, we will focus on the non-self-consistent self-energy $\pmb{\Sigma}[\mathbf{G}_0]$ in this work.

\subsection{The random phase approximation}
The GW self-energy uses some quantities from the random phase approximation (RPA), so we first define them. The RPA treats neutral electronic excitations and can be derived in several ways, including within the quasi-boson formalism discussed later. For a useful introduction, see Refs.~\citenum{rowe1968equations,ring2004nuclear}. 
\\
The RPA eigenvalue problem is defined given $\hat{h}$ and $\hat{V}$ in Eq.~\ref{eq:hamiltonian}. Assuming an orbital basis that diagonalizes $\hat{h}$, with a corresponding reference Slater determinant $|\Phi\rangle$ composed of $N$ occupied orbitals, the excitation energies $\Omega$ and amplitudes $\mathbf{X}$, $\mathbf{Y}$ are obtained from
\begin{align}
	\begin{pmatrix}
		\mathbf{A} & \mathbf{B} \\
		\mathbf{B} & \mathbf{A}
	\end{pmatrix}
	\begin{pmatrix}
		\mathbf{X} & \mathbf{Y} \\
		\mathbf{Y} & \mathbf{X}
	\end{pmatrix}
	= 
		\begin{pmatrix}
			\mathbf{X} & -\mathbf{Y} \\
			-\mathbf{Y} & \mathbf{X}
		\end{pmatrix}
	\begin{pmatrix}
		\pmb{\Omega} & \mathbf{0} \\
		\mathbf{0} & \pmb{\Omega}
	\end{pmatrix}
	\label{eq:rpa}
\end{align}
with $\mathbf{A}$, $\mathbf{B}$ matrices defined as
\begin{align}
	A_{\mu,\nu} &= A_{ia,jb} = (\epsilon_a - \epsilon_i)\delta_{ij}\delta_{ab} + V_{iajb} \label{eq:DefBMatrix}\\
	B_{\mu,\nu} &= B_{ia,jb} = V_{iabj}
	\label{eq:DefAMatrix}
\end{align}
where $i,j$ label occupied orbitals, $a,b$ label virtual orbitals, $\epsilon_i, \epsilon_a$ are the occupied and virtual eigenvalues of $\hat{h}$, 
$\Phi_i^a$, $\Phi_{ij}^{ab}$ are single and double excitations from $\Phi$. (The above corresponds to the ``direct'' RPA approximation). 
The Tamm-Dancoff approximation (TDA) to the RPA arises from setting $\mathbf{B} = 0$ (and thus $\mathbf{Y} = 0$).
$\mathbf{X}$ and $\mathbf{Y}$ are normalized such that
\begin{align}
 \begin{pmatrix}
    \mathbf{X} & \mathbf{Y} \\
    \mathbf{Y} & \mathbf{X}
    \end{pmatrix}^T
    \begin{pmatrix}\mathbf{X} & -\mathbf{Y} \\
    -\mathbf{Y} & \mathbf{X}\end{pmatrix}
    =\begin{pmatrix}
        \mathbf{1} & \mathbf{0}\\
        \mathbf{0} & \mathbf{1}
    \end{pmatrix}
\end{align}	
thus
    \begin{align}
    \begin{pmatrix}
    \mathbf{X} & \mathbf{Y} \\
    \mathbf{Y} & \mathbf{X}
\end{pmatrix}^T
    \begin{pmatrix}
    \mathbf{A} & \mathbf{B} \\
    \mathbf{B} & \mathbf{A}
\end{pmatrix}
\begin{pmatrix}
    \mathbf{X} & \mathbf{Y} \\
    \mathbf{Y} & \mathbf{X}
\end{pmatrix} = \begin{pmatrix}
    \pmb{\Omega} & \mathbf{0} \\
    \mathbf{0} & \pmb{\Omega}
\end{pmatrix}.
\end{align}
Because the RPA approximation is defined with respect to $\hat{h}$ (which is, in principle, arbitrary), it has a starting point dependence. This contributes to the starting point dependence of G$_0$W$_0$. As discussed above, we will choose $\hat{h}=\hat{f}$ in this work.

\subsection{G$_0$W$_0$, its supermatrix formulation, and GW}

The GW approximation approximates the correlation self-energy as
\begin{align}
    \Sigma^{c,\mathrm{GW}}_{pq}(\omega) = \frac{\mathrm{i}}{2\pi} \sum_{rs} \int d\omega' G_{rs}(\omega + \omega') W_{pr,qs}(\omega')
\end{align}
where $W_{pr,qs}$ are elements of the screened Coulomb interaction, itself a function $\mathbf{G}$ and $\hat{V}$. 
As discussed above, such a self-consistent self-energy must be determined iteratively, and in the first iteration, the so-called G$_0$W$_0$ version of the theory, $\mathbf{G}$ is replaced by the non-interacting Green's function and $\mathbf{W}_0$ is defined in terms of $\mathbf{G_0}$, corresponding to a treatment of screening within the RPA.
Explicitly, in the eigenvector basis of $\hat{h}$,\cite{hedin1991properties,pavlyukh2007configuration,bruneval2012ionization,van2013gw}
\begin{align}
	\Sigma^{c,\mathrm{G_0W_0}}_{pq}(\omega) =     \sum_{j\nu} \left[ W_{pj \nu} \frac{1}{\omega - \left( \epsilon_j - \Omega_\nu \right) -\mi \eta} W_{qj \nu}
	\right] + \notag  \\
	 \sum_{b\nu} \left[ W_{pb \nu} \frac{1}{\omega - \left( \epsilon_b + \Omega_\nu \right) + \mi \eta} W_{qb \nu} \right],
\label{eq:GWSelfEnergy}
\end{align}
where the screened interaction matrix element $W_{nj\nu}$ 
is defined using the RPA $\mathbf{X}$, $\mathbf{Y}$ amplitudes
\begin{align}
	W_{np \nu} = \sum_{ia}V_{npia}  \left(\mathbf{X} + \mathbf{Y}\right)_{ia,\nu}.
	\label{eq:Wh}
\end{align}
Note that $\nu$ can be interpreted as a compound index with the range of the particle-hole excitation index $ia$. 
The self-energy expression can be written compactly as
\begin{align}
    \mathbf{\Sigma}^\mathrm{c,G_0W_0}(\omega) = \mathbf{W}^T [\omega \mathbf{1} - \mathbf{d}]^{-1} \mathbf{W}
    \label{eq:matrixsigma}
    \end{align}
where $\mathbf{d}$ is the diagonal matrix with entries
\begin{align}
d_{m\nu,m\nu} = (\epsilon_m - \mathrm{sgn}(m) \Omega_\nu) 
\end{align} 
with $\mathrm{sgn}(m)=1$ if $m$ is a hole index, $\mathrm{sgn}(m)=-1$ if  $m$ is a particle index, and $\mathbf{W}$ has elements $W_{np\nu}$.
\\
Given the explicit form of the G$_0$W$_0$ correlation self-energy, we can obtain a special form of the Dyson equation where $\mathbf{G}^{\mathrm{G_0W_0}}$ is expressed as a supermatrix inversion (here supermatrix denotes a matrix of larger dimension than the Green's function). First, note for a $2 \times 2$ block matrix with block labels $1, 2$, the identity 
\begin{align}
[\mathbf{M}^{-1}]_{11} = [\mathbf{M}_{11} - \mathbf{M}_{12} \mathbf{M}_{22}^{-1} \mathbf{M}_{21}]^{-1} .
\end{align}
Then from the matrix form of the G$_0$W$_0$ self-energy,
we see that $\mathbf{G}^\mathrm{G_0W_0}$ arises from a 
$2 \times 2$ block matrix inversion
\begin{align}
\mathbf{M} &
	 = \omega \mathbf{1} - \mathbf{H}\notag\\
	 \mathbf{H}^\mathrm{G_0W_0} &= 
	\begin{pmatrix}
		\mathbf{h} + \pmb{\Sigma}^\mathrm{static} & \mathbf{W} \\
		\mathbf{W}^T & \mathbf{d}
		\end{pmatrix}\\
		\mathbf{G}^\mathrm{G_0W_0} &= [\mathbf{M}^{-1}]_{11}. \label{eq:defsupermatrix}
\end{align}
If $\mathbf{G}_0$ is defined using $\mathbf{f}$ as chosen in this work, the supermatrix simplifies to
\begin{align}
    \mathbf{H}^\mathrm{G_0W_0} &= 
	\begin{pmatrix}
		\mathbf{f} & \mathbf{W} \\
		\mathbf{W}^T & \mathbf{d}
		\end{pmatrix}
    \label{eq:hfdefsupermatrix}.
\end{align}
Since the supermatrix $\mathbf{H}$ is frequency independent, $\mathbf{G}^\mathrm{G_0W_0}$ corresponds to the resolvent of $\mathbf{H}$.
\\
The starting point dependence of $\mathrm{G_0W_0}$ enters through the choice of $\hat{h}$ (which defines $\mathbf{G}_0$ and consequently $\mathbf{W}_0$), and
can be removed once $\mathbf{G}^\mathrm{G_0W_0}$ is determined by iterating to determine a new self-energy. The fixed point of this self-consistency is the GW approximation. However, we will not discuss self-consistency in our treatment below and will focus on the $\mathrm{G}_0\mathrm{W}_0$ approximation. 
		
\section{Correspondence of G$_0$W$_0$ to the electron-boson problem}
\label{sec:EOMGW}

We next show that the supermatrix $\mathbf{H}^\mathrm{G_0W_0}$ appearing in Eq.~(\ref{eq:defsupermatrix}) arises naturally as the matrix representation of the equation-of-motion (EOM) treatment of the charged excitations of a coupled electron-boson problem.
\\
We define a coupled-electron boson Hamiltonian corresponding to a set of noninteracting electrons linearly coupled to a quadratic bath of bosons,
\begin{align}
	\hat{H}^\mathrm{eB} = \hat{H}^\mathrm{e} + \hat{H}^\mathrm{B} + \hat{V}^\mathrm{eB}  
\end{align}
where $\hat{H}^\mathrm{e}$ is the electron Hamiltonian, $\hat{H}^\mathrm{B}$ the boson Hamiltonian, and $\hat{V}^\mathrm{eB}$ is the coupling.
We choose \cite{ring2004nuclear}
\begin{align}
	\hat{H}^\mathrm{e} 
			  &= \sum_{pq} f_{pq} \hat{a}^\dagger_p \hat{a}_q\notag\\
	\hat{H}^\mathrm{B} &= \sum_{\mu,\nu}A_{\mu\nu}\hat{b}^\dagger_\mu \hat{b}_{\nu} + \frac{1}{2} \sum_{\mu,\nu}  B_{\mu\nu} \left( \hat{b}^\dagger_\mu \hat{b}^\dagger_{\nu} + \hat{b}_{\mu} \hat{b}_{\nu} \right)\notag\\
	\hat{V}^\mathrm{eB} &= \sum_{pq\nu} V_{pq\nu} \hat{a}_p^\dagger \hat{a}_q \left( \hat{b}^\dagger_\nu + \hat{b}_{\nu}\right)
\end{align}
with $\hat{b}^\dag_\nu$, $\hat{b}_\nu$ boson creation and annihilation operators, and 
\begin{align}
	V_{pq\nu} = V_{pqia}.
\end{align}
We note that $\hat{b}^\dag_\nu$, $\hat{b}_\nu$ effectively mimic fermionic particle-hole creation operators ($\hat{a}^\dagger_a \hat{a}_i \rightarrow \hat{b}^\dagger_\nu$, $\hat{a}^\dagger_i \hat{a}_a \rightarrow \hat{b}_\nu$).
However, they satisfy bosonic commutation relations and therefore violate Pauli's exclusion principle.
This is often referred to as the quasi-boson approximation. 
\\
We compute excitation energies of the coupled electron-boson Hamiltonian in the EOM formalism. Namely, given a set of operators, $\{ \hat{C}_I \}$, we solve the eigenvalue problem
\begin{align}
    \mathbf{H} \mathbf{R} = \mathbf{S} \mathbf{R} \mathbf{E}
    \label{eq:GenEigProb}
\end{align}
where the supermatrix $\mathbf{H}$ and overlap $\mathbf{S}$ are given by
\begin{align}
    H_{IJ} &= \langle 0_\mathrm{F}0_\mathrm{B}| [\hat{C}_I, [\hat{H}, \hat{C}^\dag_J]] |0_\mathrm{F}0_\mathrm{B}\rangle \notag\\
        S_{IJ} &= \langle 0_\mathrm{F}0_\mathrm{B}| [\hat{C}_I, \hat{C}^\dag_J] |0_\mathrm{F}0_\mathrm{B}\rangle \label{eq:eompropagator}
\end{align}
where $|0_\mathrm{F}\rangle = |\Phi\rangle$ and $|0_\mathrm{B}\rangle$ is the boson vacuum. 
\\
To start, we choose the operator basis $\{ \hat{C}^\dag_I \}$ = $\{ \hat{a}_i, \hat{a}_a, \hat{a}_i \hat{b}^\dag_\nu, \hat{a}_a \hat{b}_\nu\}$ (we will refer to these indices as $\{ h, p, hb, pb\}$). If we view the bosons as generating particle-hole excitations, then this basis is analogous to the standard singles and doubles bases used to describe charged excitations in quantum chemistry treatments.
Note that this set is not closed under the action of $\hat{H}^\mathrm{B}$, i.e. $[\hat{H}^\mathrm{B}, \hat{C}^\dag_I]$ can generate an operator outside of the span of $\{ \hat{C}^\dag_I\}$, because $[\hat{b}_\mu \hat{b}_\nu, \hat{b}^\dag_\nu] = \hat{b}_\mu$, but we are missing operators such as $\hat{a}_i \hat{b}_\nu, \hat{a}_a \hat{b}_\nu^\dag$ in $\{ \hat{C}^\dag_I \}$. (This is related to the problem of missing time-orderings in GW-TDA).
Evaluating the matrix elements,
we obtain $S_{IJ}=\mathrm{sgn}(I) \delta_{IJ}\ $ where $\mathrm{sgn}(I)=1$ for the hole space $\{h,hb\}$, and $\mathrm{sgn}(I)=-1$ for the particle space $\{p,pb\}$. 
The non-zero matrix elements of $\mathbf{H}$ are
\begin{align}
H_{nm} &= \mathrm{sgn}(m) F_{nm} \notag\\
H_{m,p\nu} &= \mathrm{sgn}(m) V_{m,p\nu}\notag\\
H_{n\mu,m\nu} &= \mathrm{sgn}(m) F_{nm} +  A_{\mu \nu}
\end{align}
where the sign of $H_{m,p\nu}$ depends on whether it contributes to the particle or hole space; $\mathrm{sgn}(m) = -1$ for the hole space, and $+1$ for the particle space. 
\\
Multiplying Eq.~(\ref{eq:GenEigProb}) with $-\mathbf{S}^{-1}$ from the left,  we obtain the supermatrix 
\begin{align}
	{\mathbf{H}}^\text{G$_0$W$_0$-TDA} = 
	\begin{pmatrix}
		\mathbf{F}_{h,h} & \mathbf{F}_{h,p} & \mathbf{V}_{h,hb} & \mathbf{V}_{h,pb} \\
		\mathbf{F}_{p,h} & \mathbf{F}_{p,p} & \mathbf{V}_{p,hb} & \mathbf{V}_{p,pb} \\
		\mathbf{V}^T_{{h,hb}} & \mathbf{V}^T_{{p,hb}} & \mathbf{\Delta}_{hb,hb}  & 0 \\
		\mathbf{V}^T_{{h,pb}} & \mathbf{V}^T_{{p,pb}} &  0 & \mathbf{\Delta}_{pb,pb}\\
	\end{pmatrix},
	\label{eq:DetailedMatrixform1}
\end{align}
with $\mathbf{\Delta}$ the diagonal matrix with blocks
\begin{align}
    \Delta_{hb,hb} = \mathbf{F}_{h,h} \otimes \mathbf{1}_{b,b} - \mathbf{1}_{h,h} \otimes \mathbf{A}_{bb} \notag\\ 
    \Delta_{pb,pb} = \mathbf{F}_{p,p} \otimes \mathbf{1}_{b,b} +\mathbf{1}_{p,p} \otimes \mathbf{A}_{bb}.
\end{align}
$\mathbf{H}^\textrm{G$_0$W$_0$-TDA}$ is precisely the supermatrix identified by Bintrim and Berkelbach in Ref.~\citenum{bintrim2021full}; inserting this in the Green's function definition~[Eq.~(\ref{eq:defsupermatrix})] leads to the approximation to the self-energy and Green's function termed GW-TDA in Ref.~\citenum{bintrim2021full}.
\\
The G$_0$W$_0$-TDA supermatrix is clearly similar to the G$_0$W$_0$ supermatrix but differs in the elements involving bosonic indices. For example, the diagonal matrix $\mathbf{d}$ in Eq.~\eqref{eq:defsupermatrix} is replaced by the non-diagonal energy matrix $\mathbf{\Delta}$, 
while the screened interaction $\mathbf{W}$ is replaced by the bare coupling $\mathbf{V}$.  To obtain the G$_0$W$_0$ supermatrix, we  first diagonalize the bosonic problem, i.e., rotate to the eigenbasis of $\hat{H}^\mathrm{B}$. Although such a unitary transformation does not affect the spectrum of $\hat{H}^\mathrm{eB}$, it does affect the EOM supermatrix eigenvalues when an incomplete operator basis is used. To diagonalize $\hat{H}^\mathrm{B}$, we use  a Bogoliubov transformation,\cite{ring2004nuclear}
\begin{align}
	\hat{H}^\mathrm{B}(\hat{b}, \hat{b}^\dag) = - \frac{1}{2} \mathrm{tr}\mathbf{A} + \frac{1}{2}
	\begin{pmatrix}
		\mathbf{b}^\dagger & \mathbf{b}
	\end{pmatrix}
	\begin{pmatrix}
			\mathbf{A} & \mathbf{B} \\
			\mathbf{B} & \mathbf{A}
	\end{pmatrix}
	\begin{pmatrix}
		\mathbf{b}\\
		\mathbf{b}^\dagger
	\end{pmatrix}
\end{align}
where we have used the parentheses $(\hat{b}, \hat{b}^\dag)$ to denote the operator basis in which $\hat{H}^\mathrm{B}$ is being expressed.
We recognize  
$\begin{pmatrix}
    	\mathbf{A} & \mathbf{B} \\
	    \mathbf{B} & \mathbf{A}
\end{pmatrix}$ 
as the RPA matrix in Eq.~\eqref{eq:rpa}.
We thus obtain the diagonalized form
\begin{align}
\hat{H}^\mathrm{B}(\bar{b},\bar{b}^\dag) = - \frac{1}{2} \mathrm{tr} \mathbf{A} + 
\frac{1}{2}
	\begin{pmatrix}
		\bar{\mathbf{b}}^\dagger & \bar{\mathbf{b}}
	\end{pmatrix}
		\begin{pmatrix}
			\Omega \mathbf{1} & 0 \\
			0 & \Omega \mathbf{1} 
		\end{pmatrix}
	\begin{pmatrix}
		\bar{\mathbf{b}}\\
		\bar{\mathbf{b}}^\dagger
	\end{pmatrix}
\end{align}
where 
\begin{align}
	\begin{pmatrix}
		\bar{\mathbf{b}} \\
		\bar{\mathbf{b}}^\dagger
	\end{pmatrix}
	&= 
	\begin{pmatrix}
		\mathbf{X} & -\mathbf{Y} \\
		-\mathbf{Y} & \mathbf{X}
	\end{pmatrix}^T
	\begin{pmatrix}
		\mathbf{b}\\
		\mathbf{b}^\dagger
	\end{pmatrix} \notag\\
		\begin{pmatrix}
		{\mathbf{b}} \\
		{\mathbf{b}}^\dagger
	\end{pmatrix}
	&= 
	\begin{pmatrix}
		\mathbf{X} & \mathbf{Y} \\
		\mathbf{Y} & \mathbf{X}
	\end{pmatrix}
	\begin{pmatrix}
		\bar{\mathbf{b}}\\
		\bar{\mathbf{b}}^\dagger
	\end{pmatrix}.
	\label{eq:Transformation1}
\end{align}
With respect to the new $\bar{b}$ eigenbasis, we find that 
\begin{align}
\hat{H}^\mathrm{B} &= E^c_\text{RPA} + \sum_\nu \Omega_\nu \bar{b}^\dag_\nu \bar{b}_\nu  \notag \\
	\hat{V}^\mathrm{eB} &= \sum_{pq,\nu} W_{pq,\nu} \hat{a}_p^\dagger \hat{a}_q (\bar{b}_\nu + \bar{b}^\dag_\nu)
\label{eq:linearCouplingTransformed}
\end{align}
with $E^c_\text{RPA}$ the (direct) RPA correlation energy 
\begin{align}
    E^c_\text{dRPA}= \frac{1}{2} \mathrm{tr}\{\pmb{\Omega} - \mathbf{A}\}
\end{align}
and $W_{pq,\nu}$ is the matrix element in the G$_0$W$_0$ self-energy defined in Eq.~\eqref{eq:Wh}. 
\\
We can now define the EOM supermatrix with respect to the $\{ \hat{C}^\dag_I\}$ operator basis introduced above.  Unlike in the case of G$_0$W$_0$-TDA, the operator basis is now closed under the action of $\hat{H}^\mathrm{B}$, because the non-particle number conserving bosonic terms are removed in the diagonalization. 
The non-zero matrix elements of the supermatrix are (after multiplication with $-\mathbf{S}^{-1}$)
\begin{align}
    H_{nm} &= F_{nm} \notag\\
    H_{m,p\nu} &=  W_{m,p\nu}\notag\\
    H_{n\nu,m\nu} &= F_{nm} + \mathrm{sgn}(m) \Omega_{\nu \nu}
\end{align}
and in matrix form 
\begin{align}
	{\mathbf{H}}^\mathrm{G_0W_0} = 
	\begin{pmatrix}
		\mathbf{F}_{h,h} & \mathbf{F}_{h,p} & \mathbf{W}_{h,hb} & \mathbf{W}_{h,pb} \\
		\mathbf{F}_{p,h} & \mathbf{F}_{p,p} & \mathbf{W}_{p,hb} & \mathbf{W}_{p,pb} \\
		\mathbf{W}^T_{{h,hb}} & \mathbf{W}^T_{{p,hb}} & \mathbf{d}_{hb,hb}  & 0 \\
		\mathbf{W}^T_{{h,pb}} & \mathbf{W}^T_{{p,pb}} &  0 & \mathbf{d}_{pb,pb} \\
	\end{pmatrix}.
	\label{eq:DetailedMatrixform}
\end{align}
Now consider the case starting in the eigenvector basis of $\hat{f}$. Then, $\mathbf{d}_{{hb,hb}}$, $\mathbf{d}_{{pb,pb}}$ are diagonal (and $\mathbf{F}_{{h,p}} = \mathbf{F}_{{p,h}} = 0)$. 
We see that this is identical to the G$_0$W$_0$ supermatrix defined in Eq.~\eqref{eq:hfdefsupermatrix} for the choice of $\hat{h} = \hat{f}$, showing that the $\mathrm{G}_0\mathrm{W}_0$ Green's function (starting from a Hartree-Fock $\mathbf{G}_0$)  corresponds precisely to that of an electron-boson problem in a particular excitation basis. 

\section{Insights from and relationships to equation-of-motion coupled-cluster theory}

\subsection{Background on EOM-CC and definition of the IP+EA/EOM-CC approximation}
\label{sec:BackgroundCC}

In IP/EA-EOM-CC theory, one first computes the coupled-cluster ground-state $|\Psi\rangle = e^{\hat{T}} |\Phi\rangle$
where $\hat{T}$ is the cluster excitation operator. 
At the singles and doubles level, one has $\hat{T} = \hat{T}_1 + \hat{T}_2 = \sum_{ia} t_{ia} \hat{a}^\dag_a \hat{a}_i + \frac{1}{4}\sum_{ijab} t_{ijab} \hat{a}^\dag_a \hat{a}^\dag_b \hat{a}_j \hat{a}_i$. 
Then we can define
\begin{align}
    \bar{H} = e^{-\hat{T}} \hat{H} e^{\hat{T}}  \label{eq:ccsimH}
\end{align}
where $\bar{H}$ is termed the similarity transformed Hamiltonian. Note that $\bar{H}$ is not Hermitian and thus has different left and right eigenvectors. The right ground-state is $|0_\mathrm{F}\rangle$, and we denote the left ground-state as $\langle \bar{0}_\mathrm{F}|$. The amplitudes satisfy
\begin{align}
    \langle \Phi_i^a | \bar{H}|0_\mathrm{F}\rangle &= 0 \notag \\
    \langle \Phi_{ij}^{ab} |\bar{H}|0_\mathrm{F}\rangle &= 0 \label{eq:ccamps}\\
    \langle 0_\mathrm{F} | \bar{H}|\Phi_i^a \rangle &\neq 0 \notag\\
    \langle 0_\mathrm{F} |\bar{H} |\Phi_{ij}^{ab} \rangle &\neq 0 .\label{eq:ccamps2}
\end{align}
The above means that the ground-state and neutral excitations are decoupled only in the ket space. 
\\
$\bar{H}$ can, in principle, be used in an EOM formulation as in Eq.~\ref{eq:eompropagator}. Namely, we can take the operator basis $\{\hat{C}^\dag_I \} = \{ \hat{a}_i, \hat{a}_a, \hat{a}^\dagger_a \hat{a}_i \hat{a}_j , \hat{a}^\dagger_i\hat{a}_a \hat{a}_b\}$, and
define the corresponding supermatrix elements, 
\begin{align}
\bar{H}_{IJ}^\text{IP+EA-EOM-CCSD} = \langle 0_\mathrm{F} | [\hat{C}_I, [\bar{H}, \hat{C}_J^\dag]] |0_\mathrm{F}\rangle . \label{eq:IPEAsupermatrix}
\end{align}
The standard EOM-CC formulation, however, corresponds to defining separate supermatrices for the IP and EA sectors. We do so by dividing the operator basis above into IP and EA parts, 
$\{ \hat{C}_I^{\dag \mathrm{IP}} \} =  \{ \hat{a}_i, \hat{a}^\dag_a  \hat{a}_i \hat{a}_j \}$,  
$\{ \hat{C}_I^{\mathrm{EA}} \} =  \{ \hat{a}^\dag_a, \hat{a}^\dag_a \hat{a}^\dag_b \hat{a}_j\}$. 
We then construct a supermatrix for the IP parts and EA parts separately
\begin{align}
    H_{IJ}^\text{IP-EOM-CCSD} &= \langle {0}_\mathrm{F} | [\hat{C}_I^\mathrm{IP}, [\bar{H}, \hat{C}_J^{\dag IP}]] |0_\mathrm{F} \rangle \label{eq:eablocks}\\
    H_{IJ}^\text{EA-EOM-CCSD} &= \langle {0}_\mathrm{F} | [\hat{C}_I^{\dag \mathrm{EA}}, [\bar{H}, \hat{C}_J^{\mathrm{EA}}]] |0_\mathrm{F} \rangle \label{eq:ipblocks}
\end{align}
and the overlap matrix is $S^\text{IP-EOM-CCSD}_{IJ} = \langle {0}_\mathrm{F} | [\hat{C}_I^\text{IP}, \hat{C}_J^{\dag IP}] |0_\mathrm{F} \rangle = \delta_{IJ}$ and $S^\text{EA-EOM-CCSD}_{IJ} = \langle {0}_\mathrm{F} | [\hat{C}_I^{\dag \mathrm{EA}}, \hat{C}_J^{\mathrm{EA}}] |0_\mathrm{F} \rangle = \delta_{IJ}$.
The solutions of the two eigenvalue problems then yield the IP and EA excitations. 
\\
We see there are some similarities and some differences in the construction of IP/EA-EOM-CC supermatrices and the G$_0$W$_0$ supermatrix. Of course, one difference is the replacement of the electronic Hamiltonian and excitation operators by the electron-boson Hamiltonian and electron-boson operator basis used in the quasi-boson formalism of G$_0$W$_0$. However, there are other important differences, principally the use of the similarity transformed Hamiltonian $\bar{H}$, and the neglect of the particle-hole coupling in the standard IP/EA-EOM-CC supermatrixes. To isolate the impact of these latter two choices, we can formulate an EOM-CC formalism in the quasi-boson approximation, corresponding to starting from the electron-boson Hamiltonian.
We now proceed with this analysis.

\subsection{Quasi-boson CC and exact equivalence  of equation-of-motion ring unitary coupled-cluster theory and G$_0$W$_0$}
\label{sec:BosonicDiag}

To obtain a quasi-boson formulation of EOM-CC (referred to as qb-EOM-CC in the following), we first define the ground-state CC problem as that of finding the ground-state of $\hat{H}^\mathrm{B}$. Then the CC ansatz is  
\begin{align}
    |\Psi_\mathrm{B}\rangle = e^{\hat{T}_\mathrm{B}} \ket{0_\mathrm{B}}
\end{align}
where $\hat{T}_\mathrm{B}$ is a pure boson excitation operator $\hat{T}_\mathrm{B} = \frac{1}{2} \sum_{\mu\nu} t_{\mu \nu} \hat{b}^\dagger_\mu \hat{b}^\dagger_{\nu}$. The similarity transformed boson Hamiltonian is
\begin{align}
    \bar{H}^\mathrm{B}_\mathrm{CC} &= e^{-\hat{T}_\mathrm{B}} \hat{H}^\mathrm{B} e^{\hat{T}_\mathrm{B}} \notag \\
    &= \sum_{\mu \nu} (A_{\nu \mu} + \sum_\lambda t_{\nu \lambda }B_{\lambda \mu}) \hat{b}^\dag_\nu \hat{b}_\mu + \frac{1}{2} \sum_{\mu \nu} B_{\mu \nu} \hat{b}_\mu \hat{b}_\nu
    \label{eq:STBoson}
\end{align}
and the amplitudes satisfy 
\begin{align}
\langle 0_\mathrm{B} | \hat{b}_\mu \hat{b}_\nu \bar{H}^B_\mathrm{CC} | 0_\mathrm{B} \rangle &= 0 \notag \\
\langle 0_\mathrm{B} |  \bar{H}^B_\mathrm{CC} \hat{b}^\dag_\mu \hat{b}^\dag_\nu | 0_\mathrm{B} \rangle & \neq 0. \label{eq:bosonccamplitudes}
\end{align}
It can be shown that $t_{\mu \nu} = [\mathbf{Y} \mathbf{X}^{-1}]_{\mu,\nu}$~\cite{scuseria2008ground} and using the correspondence $\nu,\mu \rightarrow ia,jb$, these amplitudes can be identified with those generated in the standard ring CC doubles (D) approximation to the amplitude equations of Eq.~(\ref{eq:ccamps})~\cite{scuseria2008ground,scuseria2013particle}. Similarly to the fermionic CC formulation,  Eq.~(\ref{eq:bosonccamplitudes}) and Eq.~(\ref{eq:STBoson}) signify that the bosonic ground-state is decoupled from the bosonic excitations only in the ket-space. 
\\
In contrast, the Bogoliubov transformation of $\hat{H}^\mathrm{B}$ in Sec.~\ref{sec:EOMGW} is a unitary (canonical) transformation. We can write the unitary operator as $e^{\hat{\sigma}_\mathrm{B}}$ with $\hat{\sigma} = \frac{1}{2} \sum_{\mu\nu} \sigma_{\mu \nu} (\hat{b}^\dag_\mu \hat{b}^\dag_\nu - \hat{b}_\mu \hat{b}_\nu)$,   $\hat{\sigma}^\dag_\mathrm{B} =-\hat{\sigma}_\mathrm{B}$. This leads to a bosonic version of the {unitary} coupled-cluster (uCC) ground-state theory,
\begin{align}
|\Psi^\mathrm{B}_\mathrm{uCC}\rangle 
	= e^{\hat{\sigma}_\mathrm{B}} \ket{0_\mathrm{B}} \equiv \ket{\bar{0}_\mathrm{B}}. \label{eq:ctbosongs}
\end{align}
The corresponding bosonic unitary transformed Hamiltonian $\bar{H}^\mathrm{B}_\mathrm{uCC}$ and amplitude equations satisfy  
\begin{align}
    \langle \mu \nu | \bar{H}^\mathrm{B}_\mathrm{uCC} | 0_\mathrm{B}\rangle &= \langle 0_\mathrm{B} | \bar{H}^\mathrm{B}_\mathrm{uCC} | \mu \nu \rangle = 0 \label{eq:uccamps}
\end{align}
and
\begin{align}
    \bar{H}^\mathrm{B}_\mathrm{uCC} = \sum_{\mu\nu} \bar{A}_{\mu \nu} \hat{b}^\dag_\mu \hat{b}_\nu. \label{eq:ctbosonH}
\end{align}
In the bosonic uCC, the  bosonic excitations are decoupled from the ground-state in both the bra and ket spaces, and  we can further choose $\hat{\sigma}_\mathrm{B}$ such that the Hamiltonian is fully diagonalized, as we did in the analysis of G$_0$W$_0$ above.
The bosonic uCC amplitude equations correspond to a direct ring-version of the standard uCCD amplitude equations. 
\\
Although the bosonic CC and bosonic uCC theories create the same ground-state of $\hat{H}^\mathrm{B}$, i.e. the RPA ground-state, up to a normalization factor $\mathit{N}_0$,
\begin{align}
|\Psi_\mathrm{uCC}^\mathrm{B}\rangle = \mathit{N}_0 |\Psi_\mathrm{CC}^\mathrm{B}\rangle
\label{eq:Thouless}
\end{align}
the different form of $\bar{H}^\mathrm{B}_\mathrm{CC}$ and $\bar{H}^\mathrm{B}_\mathrm{uCC}$ lead to different EOM approximations. To see this, we first extend the above analysis to the full electron-boson Hamiltonian $\hat{H}^\mathrm{eB}$. This is simple because if $|\Psi_\mathrm{CC}^\mathrm{B}\rangle$,  $|\Psi_\mathrm{uCC}^\mathrm{B}\rangle$ are eigenstates of $\hat{H}^\mathrm{B}$ then $|0_\mathrm{F} \Psi_\mathrm{CC}^\mathrm{B} \rangle$, $|0_\mathrm{F} \Psi_\mathrm{uCC}^\mathrm{B} \rangle$
are eigenstates of $\hat{H}^\mathrm{eB}$ with the same eigenvalue. We can similarly define the similarity and unitarily transformed electron-boson Hamiltonians
\begin{align}
    \bar{H}^\mathrm{eB}_\mathrm{CC} = e^{-\hat{T}_\mathrm{B}} \hat{H}^\mathrm{eB} e^{\hat{T}_\mathrm{B}} \label{eq:STBCC} \\
    \bar{H}^\mathrm{eB}_\mathrm{uCC} = e^{-\hat{\sigma}_\mathrm{B}} \hat{H}^\mathrm{eB} e^{\hat{\sigma}_\mathrm{B}}.\label{eq:UTBCC}
\end{align}
Then $\bar{H}^\mathrm{eB}_\mathrm{uCC}$ is identical to the Bogoliubov transformed Hamiltonian of the above analysis of $\mathrm{G}_0\mathrm{W}_0$ theory. Consequently, within the electron-boson picture the IP+EA-qb-EOM-uCCD  is precisely equivalent to $\mathrm{G}_0\mathrm{W}_0$ theory.
\\
In contrast, if we define an IP+EA-qb-EOM formulation using the similarity transformed electron-boson Hamiltonian $\bar{H}^\mathrm{eB}_\mathrm{CC}$ (denoted as IP+EA-qb-EOM-CCD), the  non-number conserving nature of $\bar{H}^\mathrm{B}_\mathrm{CC}$ means that the $\mathrm{G}_0\mathrm{W}_0$ operator basis is not closed under the action of $\bar{H}^\mathrm{B}_\mathrm{CC}$.
Thus, although some non-TDA screening contributions are included in this treatment, others are not, and this may be viewed as an expression of the missing time-orderings in the screening terms in IP/EA-EOM-CCSD which was diagrammatically analyzed in Ref.~\citenum{lange2018relation}. We examine the numerical consequences of this in Sec.~\ref{sec:ccvsucc}.

\subsection{Decoupling of particle and hole subspaces}
\label{sec:phDecoupling}
A second source of differences between the $\mathrm{G}_0\mathrm{W}_0$ approximation and the IP/EA-EOM-CC treatments
arises from the neglect of the IP and EA coupling in the IP/EA-EOM-CC supermatrices.  
Reordering the $\mathbf{H}^\mathrm{G_0W_0}$ supermatrix, we write
\begin{align}
	{\mathbf{H}}^\mathrm{G_0W_0} &= 
	\begin{pmatrix}
		\mathbf{F}_{h,h} & \mathbf{W}_{h,hb} & \mathbf{F}_{h,p} & \mathbf{W}_{h,pb} \\
		\mathbf{W}_{{hb,h}} & \mathbf{d}_{hb,hb} & \mathbf{W}_{{hb,p}} & \mathbf{0} \\
		\mathbf{F}_{p,h}  & \mathbf{W}_{p,hb}& \mathbf{F}_{p,p} & \mathbf{W}_{p,pb}\\
		\mathbf{W}_{pb,h} & \mathbf{0} & \mathbf{W}_{pb,p} & \mathbf{d}_{pb,pb}\\
		\end{pmatrix} \notag \\
		&=\begin{pmatrix}
		\mathbf{H}^\mathrm{IP} & \mathbf{C}^\mathrm{IP/EA} \\
		 \mathbf{C}^\mathrm{EA/IP} & \mathbf{H}^\mathrm{EA}\\  
	\end{pmatrix},
	\label{eq:DetailedMatrixformReordered}
\end{align}
and we see that $\mathbf{C}$ is the coupling between the IP and EA sectors. Analogous to in the IP/EA-EOM-CC treatment we could choose to neglect the coupling matrix, and diagonalize the IP and EA blocks of the supermatrix separately. We will term this the IP/EA-G$_0$W$_0$ approximation. 
\\
This, however, is not a perfect analogy, because if the ground-state CC amplitude equations are satisfied up to some level of excitation, then some elements of $\mathbf{C}$ are removed in the IP+EA-EOM-CC supermatrix. (This has been discussed in the context of unitary coupled-cluster theory in  Ref.~\citenum{mukherjee1989effective}). For example, if $\hat{C}^{\dag \mathrm{EA}} = \hat{a}_a$ and $\hat{C}^{\mathrm{IP}} = \hat{a}^\dag_i \hat{a}^\dag_j \hat{a}_b  $, then 
\begin{align}
    \langle 0_\mathrm{F}| [\hat{C}^\mathrm{IP}, [\bar{H}, \hat{C}^{\dag \mathrm{EA}}]] |0_\mathrm{F}\rangle 
    =
    -\langle 0_\mathrm{F} | \hat{a}^\dag_i \hat{a}^\dag_j \hat{a}_b \hat{a}_a  \bar{H} | 0_\mathrm{F} \rangle
\end{align} 
and if $\bar{H}$ is constructed with CCSD amplitudes, this term is $0$ [see Eq.~\eqref{eq:ccamps}]. In other words, the ground-state amplitude equations mean that the IP+EA-EOM-CCSD supermatrix is (partially) decoupled between the particle and hole sectors, and IP/EA-EOM-CCSD approximations only neglect the residual coupling.  
Identifying boson labels in $\{C^\dagger_I\}$ with the excitations $ia$,
the supermatrix in Eq.~\eqref{eq:IPEAsupermatrix}
has the structure
\begin{align}
	{\mathbf{H}}^\textrm{IP+EA-EOM-CCSD} = 
	\begin{pmatrix}
		\mathbf{F}_{h,h} & \mathbf{W}_{h,hb} & \mathbf{0} & \mathbf{0} \\
		\mathbf{W}_{{hb,h}} & \mathbf{D}_{hb,hb} & \mathbf{0} & \mathbf{C}_{hb,pb} \\
		\mathbf{F}_{p,h}  & \mathbf{W}_{{p,hb}} & \mathbf{F}_{p,p} & \mathbf{W}_{p,pb}\\
		\mathbf{W}_{{pb,h}} & \mathbf{0} & \mathbf{W}_{{pb,p}} & \mathbf{D}_{pb,pb}\\
	\end{pmatrix} . \label{eq:ccsuper}
\end{align}
(For notational economy in this section, we reuse the symbols $\mathbf{F}$, $\mathbf{W}$, etc., even though their values are different in different supermatrices.  
$\mathbf{D}$ is in general non-diagonal). 
IP/EA-EOM-CCSD corresponds to neglecting only the $\mathbf{C}_{hb, pb}$ coupling (in this case the lower $2\times 2$ block of couplings does not affect the eigenvalues of the supermatrix, although it does affect the left eigenvectors).
\\
To understand the quality of this approximate decoupling, 
it is instructive to pursue a similar decoupling of the electron-boson Hamiltonian in the $\mathrm{G}_0\mathrm{W}_0$ setting. Here we assume that we have already performed the unitary Bogoliubov transformation to diagonalize $\hat{H}^\mathrm{B}$, i.e. we are using $\bar{H}^\mathrm{eB}$,
whose supermatrix is $\mathbf{H}^\mathrm{G_0W_0}$.
As established, this is equivalent to $\mathbf{H}^\textrm{IP+EA-qb-EOM-uCCD}$ in the electron-boson setting, but we will primarily view this as a decoupling of the $\mathrm{G}_0\mathrm{W}_0$ theory in this section. 
We can then require that our correlated vacuum $|0_\textrm{F}\bar{0}_\textrm{B}\rangle$ satisfies  amplitude equations which set parts of the coupling matrix $\mathbf{C}$ to zero. In the standard CC theory, such amplitude equations are the same as the ground-state amplitude equations, but appear as additional amplitude equations in the current setting.  
We can decouple only the upper-block interactions, similar to IP/EA-EOM-CC theory, using a pure excitation operator $\hat{T}^{\mathrm{eB}}$ and a similarity transform
\begin{align}
    \hat{T}^{\mathrm{eB}} &= \hat{T}^{\mathrm{eB}}_1 + \hat{T}^{\mathrm{eB}}_2 \nonumber\\
    &=\sum_{ia} t_{ia} \hat{a}^\dag_a \hat{a}_i + \sum_{\mu ia} t_{ia\mu} \hat{a}^\dag_a \hat{a}_i \bar{b}^\dag_\mu \label{eq:Toperator} \\\
      0 &= \langle 0_\mathrm{F} \bar{0}_\mathrm{B}|\hat{a}^\dagger_i \hat{a}_a e^{-\hat{T}^{\mathrm{eB}}} {\bar{H}}^\mathrm{eB} e^{\hat{T}^{\mathrm{eB}}} |0_\mathrm{F} \bar{0}_\mathrm{B}\rangle \notag \\
     0 &= \langle 0_\mathrm{F} \bar{0}_\mathrm{B}|\hat{a}^\dagger_i \hat{a}_a \bar{b}_\mu e^{-\hat{T}^{\mathrm{eB}}} {\bar{H}}^\mathrm{eB} e^{\hat{T}^{\mathrm{eB}}} |0_\mathrm{F} \bar{0}_\mathrm{B}\rangle
\end{align}
yielding the ST-$\mathrm{G}_0\mathrm{W}_0$ supermatrix
\begin{align}
	{\mathbf{H}}^\mathrm{ST-G_0W_0} = 
	\begin{pmatrix}
		\mathbf{F}_{h,h} & \mathbf{W}_{h,hb} & \mathbf{0} & \mathbf{0} \\
		\mathbf{W}_{\mathrm{hb,h}} & \mathbf{D}_{hb,hb} & \mathbf{0} & \mathbf{C}_{hb,pb} \\
		\mathbf{F}_{p,h}  & \mathbf{W}_{{p,hb}} & \mathbf{F}_{p,p} & \mathbf{W}_{p,pb}\\
		\mathbf{W}_{{pb,h}} & \mathbf{0} & \mathbf{W}_{{pb,p}} & \mathbf{D}_{pb,pb}\\
	\end{pmatrix} . \label{eq:STDec}
\end{align}
Diagonalizing the IP and EA blocks (evaluated in the $\{\hat{C}^{\dag \mathrm{IP}}_I\}$ and $\{\hat{C}^{ \mathrm{EA}}_I\}$ basis) separately yields what we term the IP/EA-ST-G$_0$W$_0$ approximations.
Alternatively, we can use a unitary operator $\hat{\sigma}^{\mathrm{eB}} = \hat{T}^{\mathrm{eB}} - \hat{T}^{\mathrm{eB}\dag}$ and require 
\begin{align}
    0&=\langle 0_\mathrm{F} \bar{0}_\mathrm{B}| [e^{-{\hat{\sigma}^\mathrm{eB}}} \bar{H}^\mathrm{eB} e^{\hat{\sigma}^\mathrm{eB}}, \hat{a}^\dag_a \hat{a}_i - \hat{a}^\dag_i \hat{a}_a ] |0_\mathrm{F} \bar{0}_\mathrm{B}\rangle \notag\\ 
    0&=\langle 0_\mathrm{F} \bar{0}_\mathrm{B}| [e^{-{\hat{\sigma}^\mathrm{eB}}} \bar{H}^\mathrm{eB} e^{\hat{\sigma}^\mathrm{eB}}, \hat{a}^\dag_a \hat{a}_i \bar{b}^\dag_\mu - \hat{a}^\dag_i \hat{a}_a \bar{b}_\mu ] |0_\mathrm{F} \bar{0}_\mathrm{B}\rangle   
\end{align}
leading to the symmetrical block structure
 \begin{align}
 	{\mathbf{H}}^\mathrm{UT-G_0W_0} = 
	\begin{pmatrix}
		\mathbf{F}_{h,h} & \mathbf{W}_{h,hb} & \mathbf{0} & \mathbf{0} \\
		\mathbf{W}_{{hb,h}} & \mathbf{D}_{hb,hb} & \mathbf{0} & \mathbf{C}_{hb,pb} \\
		\mathbf{0}  & \mathbf{0} & \mathbf{F}_{p,p} & \mathbf{W}_{p,pb}\\
		\mathbf{0} & \mathbf{C}_{pb,hb} & \mathbf{W}_{{pb,p}} & \mathbf{D}_{pb,pb}\\
	\end{pmatrix} . \label{eq:UTDec}
 \end{align}
Diagonalizing the IP and EA blocks separately leads then to the IP/EA-UT-G$_0$W$_0$ approximations.
\\
The decouplings above do not correspond to 
a simple matrix decoupling of the Hamiltonian, 
because the transformations are done at the second-quantized level (i.e. in terms of infinite matrices) before being projected down into the finite operator basis. 
Consequently, the eigenvalues of the G$_0$W$_0$, UT-G$_0$W$_0$, ST-G$_0$W$_0$ supermatrices
are all different from each other. 
Because we are using second-quantized transformations, the particle-hole decoupled Hamiltonians 
contain different interactions from $\bar{H}^{\mathrm{eB}}$ even in the purely electronic sector. This is a type of renormalization of the electronic Green's function. In the Baker-Campbell-Hausdorff  (BCH) expansion of these Hamiltonians, the exact expansion of the commutators leads to higher particle  electronic interactions. 
To control the complexity, we use an approximate BCH expansion that retains the same interaction form as the original $\bar{H}^\mathrm{eB}$ Hamiltonian, 
\begin{align}
    \bar{H}^\mathrm{eB}_\textrm{decoupled(ST)} &= \bar{H}^{\mathrm{eB}} + [\bar{H}^{\mathrm{eB}}, \hat{T}^{\mathrm{eB}}]_{1} \notag \\ &+ \frac{1}{2} [[\bar{H}^{\mathrm{eB}},\hat{T}^{\mathrm{eB}}], \hat{T}^\mathrm{eB}]_1 + \ldots \label{eq:decoupledST} \\
    \bar{H}^\mathrm{eB}_\textrm{decoupled(UT)} &= \bar{H}^{\mathrm{eB}} + [\bar{H}^{\mathrm{eB}}, \hat{\sigma}^{\mathrm{eB}}]_{1} \notag \\ &+ \frac{1}{2} [[\bar{H}^{\mathrm{eB}},\hat{\sigma}^{\mathrm{eB}}]_1, \hat{\sigma}^{\mathrm{eB}}]_1 + \dots \label{eq:decoupledUT}
\end{align}
where the subscript $1$ indicates that only up to 1-particle electronic terms after normal ordering with respect to  $|0_\mathrm{F}\rangle$ are retained. This approximation includes mean-field-like contributions of the 2- and higher-particle electronic interactions generated in the decoupling transformation. The BCH expansion for the similarity transformed (ST) Hamiltonian naturally truncates at third order and 2-particle interactions. Because the $\mathrm{G}_0\mathrm{W}_0$ operator basis does not contain more than single electron holes or particles, the terms dropped in the ST expansion do not  contribute to the ST-$\mathrm{G}_0\mathrm{W}_0$ supermatrix. 
Numerical results for the particle-hole decoupling transformations  are presented in 
Sec.~\ref{sec:phDecouplingNumerical}.

\section{Numerical investigations}

We now describe numerical experiments to illustrate our above analysis, in particular, with respect to the   different treatments of screening in G$_0$W$_0$, the relationship between G$_0$W$_0$ and EOM-CC theories, and the effectiveness of approximate particle-hole decoupling. Each experiment corresponds to constructing a different Hamiltonian supermatrix, which we diagonalize to report the poles, focusing mainly on the quasiparticle HOMO and LUMO energies. 
In the case of G$_0$W$_0$, this means that the quasiparticle poles reported include the  contributions of the full self-energy matrix, i.e.~we do not employ the common diagonal approximation (similarly to Refs.~\citenum{kaplan2015off,bintrim2021full}).
\\
The working equations for all methods were generated using \textsc{wick} \cite{wick} and the methods were implemented using  \textsc{PySCF}.\cite{sun2018pyscf,sun2020recent}
The ground-state direct ring CCD amplitudes appearing in IP+EA-qb-EOM-CCD  obtained from the RPA $\mathbf{X}$, $\mathbf{Y}$ matrices, while the direct ring unitary CC doubles Hamiltonian ($\bar{H}^\mathrm{eB}_\mathrm{uCC}$) is constructed based on canonical transformation theory~\cite{yanai2006canonical,neuscamman2010review}. The required commutator expressions are given in Appendix \ref{sec:CT-RPA}.
The IP/EA decoupling amplitude equations in the similarity transformed approach are given explicitly in Appendix~\ref{sec:amplitudequations}, while the unitary transformation decoupling equations were solved within the canonical transformation approach. 
A list of abbreviations for the variety of EOM approaches is provided in the Appendix \ref{sec:abbreviations}.
All molecular structures were taken from the GW100 testset\cite{van2015gw} and we used the def2-TZVP \cite{weigend2005balanced} basis in all calculations. All calculations used a Hartree-Fock Green's function as $\mathbf{G}_0$.

\subsection{G$_0$W$_0$,  G$_0$W$_0$-TDA, IP+EA-qb-EOM-(u)CCD}
\label{sec:ccvsucc}
We first investigate the effect of different treatments of RPA screening on the quasiparticle energies given by G$_0$W$_0$, 
G$_0$W$_0$-TDA, and the two EOM-CC theories for the electron-boson Hamiltonian, denoted IP+EA-qb-EOM-CCD and IP+EA-qb-EOM-uCCD. 
We show the G$_0$W$_0$ HOMO/LUMO quasiparticle energies, fundamental gap $E^\mathrm{Gap}$, and the differences from G$_0$W$_0$ (reported as mean absolute errors (MAE))  in Tab.~\ref{tab:RPAScreening}.
\\
The differences between G$_0$W$_0$-TDA and G$_0$W$_0$ capture the contribution of non-TDA screening to the quasiparticle energies.
We see that the neglect of non-TDA screening results in a more positive HOMO  and a less positive LUMO energy (with a few exceptions) and consequently a smaller fundamental gap.
The non-TDA terms have a greater effect on the HOMO than the LUMO quasiparticle energies. We  observe a MAE for G$_0$W$_0$-TDA of 0.336~eV (HOMO) and 0.081~eV (LUMO) and the MAE for $\Delta E^\mathrm{Gap}$ is $0.390$~eV.
\\
IP+EA-qb-EOM-CCD 
includes a subset of the non-TDA screening contributions and 
this results in a decrease of the MAE for the HOMO and LUMO quasiparticle energies to $0.191$~eV 
and $0.041$~eV, and $0.218$~eV for $\Delta E^\mathrm{Gap}$.
However, deviations from G$_0$W$_0$ of more than $0.5$ eV (HOMO, MgO) are still observed, highlighting the inability of the similarity transformation to capture some important effects of RPA screening.
\\
IP+EA-qb-EOM-uCCD yields numerically identical quasiparticle energies to G$_0$W$_0$.
This demonstrates the equivalence of the bosonic unitary transformation in IP+EA-qb-EOM-uCCD to RPA screening, as already shown theoretically in Sec.~\ref{sec:BosonicDiag}.
\\
In Tab.~\ref{tab:RPAScreening2}, we show more detailed data on the quasiparticles energies from HOMO-2 to LUMO+2 for F$_2$ and MgO. From the deviations from G$_0$W$_0$ it is clear that neglecting parts of RPA screening does not result in a uniform shift in the quasiparticle IPs or EAs, e.g.~the deviations for G$_0$W$_0$-TDA for the HOMO and HOMO-1 of MgO are $0.968$ eV and $0.203$ eV, respectively.
Interestingly, we see particularly large deviations for the HOMO-2 of MgO. This appears to be related to the overall small quasiparticle weight (i.e. the total norm in the single-particle sector for excitation $n$, $||\mathbf{R}^n_h||^2 + ||\mathbf{R}^n_p||^2$) of $0.31$ (G$_0$W$_0$-RPA), $0.54$ (G$_0$W$_0$-TDA), $0.39$ (IP+EA-qb-EOM-CCD). 
This means that there are large contributions from the bosonic sectors of the Hamiltonian supermatrix, which is the part that is treated differently in all the approaches. 
\begin{table*}
        \scriptsize
		\centering
		\caption{HOMO/LUMO quasiparticle energies and fundamental gap $E^\mathrm{Gap}$ of G$_0$W$_0$@HF for various molecules and differences from G$_0$W$_0$@HF in eV (G$_0$W$_0$-TDA: G$_0$W$_0$ within the Tamm--Dancoff Approximation [Eq.~(\ref{eq:DetailedMatrixform1})], IP+EA-qb-EOM-CCD: Similarity-transformed electron-boson Hamiltonian [Eq.~(\ref{eq:STBCC})],
		IP+EA-qb-EOM-uCCD: Unitary-transformed electron-boson Hamiltonian [Eq.~(\ref{eq:UTBCC})] for EOM, MAE: Mean Absolute Error (deviation from G$_0$W$_0$@HF), def2-TZVP).}
		\label{tab:RPAScreening}
		\begin{tabular}{l | c c c | c c c | c c c | c c c} 
\hline
 & \multicolumn{3}{c}{G$_0$W$_0$} & \multicolumn{3}{c}{G$_0$W$_0$-TDA} & \multicolumn{3}{c}{IP+EA-qb-EOM-CCD} & \multicolumn{3}{c}{IP+EA-qb-EOM-uCCD} \\
\hline
Molecule & HOMO & LUMO & $E^\mathrm{Gap}$ & $\Delta$HOMO & $\Delta$LUMO & $\Delta$$E^\mathrm{Gap}$ & $\Delta$HOMO & $\Delta$LUMO & $\Delta$$E^\mathrm{Gap}$ & $\Delta$HOMO & $\Delta$LUMO & $\Delta$$E^\mathrm{Gap}$  \\
\hline
He & $-$24.301 & 22.401 & 46.702 & 0.143 & $-$0.025 & $-$0.168 & 0.076 & $-$0.013 & $-$0.089  & 0.000 & 0.000 & 0.000 \\
Ne & $-$21.362 & 21.197 & 42.559 & 0.605 & $-$0.077 & $-$0.682 & 0.332 & $-$0.040 & $-$0.372  & 0.000 & 0.000 & 0.000 \\
H$_2$ & $-$16.308 & 4.404 & 20.712 & $-$0.027 & $-$0.006 & 0.021 & $-$0.009 & $-$0.003 & 0.006 & 0.000 & 0.000 & 0.000 \\
Li$_2$ & $-$5.165 & 0.018 & 5.183 & $-$0.056 & $-$0.068 & $-$0.012 & $-$0.024 & $-$0.034 & $-$0.010 & 0.000 & 0.000 & 0.000 \\
F$_2$ & $-$16.274 & 0.753 & 17.027 & 0.790 & $-$0.208 & $-$0.998 & 0.431 & $-$0.106 & $-$0.537 & 0.000 & 0.000 & 0.000 \\
SiH$_4$ & $-$13.082 & 3.341 & 16.423 & 0.055 & $-$0.107 & $-$0.162 & 0.034 & $-$0.053 & $-$0.087 & 0.000 & 0.000 & 0.000 \\
LiH & $-$7.949 & 0.123 & 8.072 & 0.112 & $-$0.009 & $-$0.121 & 0.062 & $-$0.004 & $-$0.066 & 0.000 & 0.000 & 0.000 \\
CO & $-$14.99 & 1.094 & 16.084 & 0.220 & $-$0.087 & $-$0.307 & 0.131 & $-$0.042 & $-$0.173  & 0.000 & 0.000 & 0.000 \\
H$_2$O & $-$12.789 & 3.114 & 15.903 & 0.464 & $-$0.058 & $-$0.522 & 0.265 & $-$0.029 & $-$0.294  & 0.000 & 0.000 & 0.000 \\
BeO & $-$9.788 & $-$2.097 & 7.691 & 0.366 & $-$0.050 & $-$0.416 & 0.233 & $-$0.026 & $-$0.259 & 0.000 & 0.000 & 0.000 \\
MgO & $-$7.863 & $-$1.506 & 6.357 & 0.968 & 0.132 & $-$0.836 & 0.562 & 0.071 & $-$0.491 & 0.000 & 0.000 & 0.000 \\
H$_2$CO & $-$11.206 & 1.822 & 13.028 & 0.446 & $-$0.191 & $-$0.637 & 0.249 & $-$0.094 & $-$0.343 & 0.000 & 0.000 & 0.000 \\
CH$_4$ & $-$14.637 & 3.650 & 18.287 & 0.102 & $-$0.076 & $-$0.178 & 0.064 & $-$0.038 & $-$0.102 & 0.000 & 0.000 & 0.000 \\
SO$_2$ & $-$12.827 & $-$0.483 & 12.344 & 0.353 & $-$0.045 & $-$0.398 & 0.203 & $-$0.020 & $-$0.223 & 0.000 & 0.000 & 0.000 \\
\hline 
MAE & & & & 0.336 & 0.081 & 0.390 & 0.191 & 0.041 & 0.218 & 0.000 & 0.000 & 0.000 \\
\hline
	\end{tabular}
\end{table*}

\begin{table*}
        \scriptsize
		\centering
		\caption{Quasiparticle energies of G$_0$W$_0$@HF and differences from G$_0$W$_0$@HF in eV (G$_0$W$_0$-TDA: G$_0$W$_0$ within the Tamm--Dancoff Approximation [Eq.~(\ref{eq:DetailedMatrixform1})], IP+EA-qb-EOM-CCD: Similarity-transformed electron-boson Hamiltonian [Eq.~(\ref{eq:STBCC})] for EOM, the degeneracy of the quasiparticle energies is shown in parentheses, def2-TZVP).}
		\label{tab:RPAScreening2}
		\begin{tabular}{l | c c c | c c c } 
\hline
& \multicolumn{3}{c|}{F$_2$} & \multicolumn{3}{c}{MgO} \\
\hline
& G$_0$W$_0$ & $\Delta$G$_0$W$_0$-TDA & $\Delta$IP+EA-qb-EOM-CCD &  G$_0$W$_0$ & $\Delta$G$_0$W$_0$-TDA & $\Delta$IP+EA-qb-EOM-CCD \\
\hline
HOMO-2 & $-$20.773	    &  $-$0.267    & $-$0.106  & $-$25.309& 2.567    & 2.058\\
HOMO-1 & $-$19.863(2x) &  0.896(2x)  & 0.501(2x) & $-$8.444 & 0.203    & 0.157\\ 
HOMO   & $-$16.274(2x) &  0.790(2x)& 0.431(2x) & $-$7.863(2x)  & 0.968(2x)& 0.562(2x)\\
LUMO   & 0.753          & $-$0.208     &  $-$0.106 & $-$1.506 & 0.132      & 0.071 \\
LUMO+1 & 15.778         & $-$0.270     & $-$0.120  & 1.088(2x) & $-$0.026(2x) &$-$0.015(2x)\\
LUMO+2 & 15.828         & $-$0.254     & $-$0.157  & 2.606 & $-$0.092 &   $-$0.053  \\
\hline
	\end{tabular}
\end{table*}

\subsection{Particle-hole decoupling}
\label{sec:phDecouplingNumerical}

We now investigate the approximate particle-hole (IP/EA) decoupling of G$_0$W$_0$ (inspired by the analogous particle-hole decoupling of EOM-CC) as discussed in Sec.~\ref{sec:phDecoupling}.  
The MAE deviations from G$_0$W$_0$ of the HOMO and LUMO quasiparticle energies for the  IP/EA-G$_0$W$_0$ (neglecting any IP/EA coupling), similarity transformed decoupled IP/EA-ST-G$_0$W$_0$ [Eq.~(\ref{eq:STDec})] and unitary transformed decoupled IP/EA-UT-G$_0$W$_0$ [Eq.~(\ref{eq:UTDec})] are shown in Tab.~\ref{tab:Decoupling1}.
\\
Simply neglecting particle-hole couplings as in IP/EA-G$_0$W$_0$ gives a MAE of $1.579$ eV for the HOMO and $0.578$ eV for the LUMO quasiparticle energies.
These large deviations highlight the importance of particle-hole coupling in G$_0$W$_0$ (similar results are obtained for G$_0$W$_0$-TDA in Ref.~\citenum{bintrim2021full}).
The approximate decoupling procedures, IP/EA-ST-G$_0$W$_0$ and IP/EA-UT-G$_0$W$_0$, significantly reduce the deviations to a MAE of $0.061$ eV (HOMO) / $0.027$ eV (LUMO) and $0.061$ eV (HOMO) / $0.019$ eV (LUMO), respectively.
The remaining deviation is due to the approximate nature of the decoupling from the truncation of the cluster expansion at the singles/doubles level and (in the case of the unitary transform) the neglect of some of the new electronic interactions generated by the unitary transformation (Sec.~\ref{sec:phDecoupling}).
\\
Deviations from G$_0$W$_0$ for the HOMO-2 to LUMO+2 quasiparticle energies are shown for  F$_2$ and MgO in Tab.~\ref{tab:Decoupling2}. 
We see that the particle-hole coupling contributions vary significantly depending on the quasiparticle, e.g.~in IP/EA-G$_0$W$_0$ the deviations for the LUMO and LUMO+1 of F$_2$ are $-2.120$ eV and $-0.355$ eV, respectively; 
for IP/EA-ST-G$_0$W$_0$ and IP/EA-UT-G$_0$W$_0$ the dependency is less pronounced.
We see large deviations  for the HOMO-2 of MgO ($2.309$~eV IP/EA-G$_0$W$_0$, $1.174$~eV IP/EA-ST-G$_0$W$_0$, and $1.049$~eV IP/EA-UT-G$_0$W$_0$).
As already discussed in Sec.~\ref{sec:ccvsucc}, the quasiparticle weight in this case is small, highlighting a strong interaction of the electron/hole with the quasi-bosons. 
\begin{table}
        \scriptsize
		\centering
		\caption{HOMO/LUMO quasiparticle energy differences from G$_0$W$_0$@HF in eV (IP/EA-G$_0$W$_0$: G$_0$W$_0$ calculation excluding particle-hole coupling, IP/EA-ST-G$_0$W$_0$: Singles-doubles similarity transformation (ST) for particle hole decoupling, IP/EA-UT-G$_0$W$_0$: Singles-doubles unitary transformation (UT) for particle hole decoupling, MAE: Mean Absolute Error (deviation from G$_0$W$_0$@HF), def2-TZVP).}
		\label{tab:Decoupling1}
		\begin{tabular}{l | c c | c c | c c}
            \hline
             & \multicolumn{2}{c}{IP/EA-G$_0$W$_0$} & \multicolumn{2}{c}{IP/EA-ST-G$_0$W$_0$} & \multicolumn{2}{c}{IP/EA-UT-G$_0$W$_0$}\\
             \hline
            Molecule & $\Delta$HOMO & $\Delta$LUMO & $\Delta$HOMO & $\Delta$LUMO &  $\Delta$HOMO & $\Delta$LUMO \\
            \hline
He & 1.178 & $-$0.333 & $-$0.007 & $-$0.002 & 0.000 & $-$0.006 \\
Ne & 2.059 & $-$0.375 & 0.077 & $-$0.004 & 0.086 & $-$0.013 \\
H$_2$ & 1.162 & $-$0.125 & $-$0.023 & $-$0.004 & $-$0.016 & $-$0.002 \\
Li$_2$ & 0.598 & $-$0.094 & $-$0.008 & $-$0.007 & $-$0.016 & $-$0.005 \\
F$_2$ & 2.083 & $-$2.120 & 0.094 & $-$0.091 & 0.134 & $-$0.067 \\
SiH$_4$ & 1.257 & $-$0.207 & $-$0.020 & $-$0.016 & $-$0.017 & $-$0.011 \\
LiH & 0.845 & $-$0.048 & $-$0.026 & $-$0.002 & $-$0.002 & $-$0.001 \\
CO & 1.803 & $-$1.011 & $-$0.024 & $-$0.063 & 0.039 & $-$0.023 \\
H$_2$O & 1.866 & $-$0.216 & 0.040 & $-$0.009 & 0.090 & $-$0.008 \\
BeO & 1.920 & $-$0.131 & 0.090 & $-$0.020 & 0.097 & $-$0.004 \\
MgO & 1.845 & $-$0.801 & 0.341 & $-$0.006 & 0.222 & $-$0.029 \\
H$_2$CO & 1.904 & $-$0.999 & 0.052 & $-$0.066 & 0.065 & $-$0.043 \\
CH$_4$ & 1.549 & $-$0.15 & $-$0.026 & $-$0.013 & 0.005 & $-$0.009 \\
SO$_2$ & 2.032 & $-$1.483 & 0.028 & $-$0.076 & 0.060 & $-$0.042 \\
            \hline
            MAE & 1.579 & 0.578 & 0.061 & 0.027 & 0.061 & 0.019 \\
            \hline
		\end{tabular}
  \label{tab:PH}
\end{table}
\begin{table*}
        \scriptsize
		\centering
		\caption{Quasiparticle energy differences from G$_0$W$_0$@HF in eV (IP/EA-G$_0$W$_0$: G$_0$W$_0$ calculation with exclusion of particle-hole coupling, IP/EA-ST-G$_0$W$_0$: Singles-doubles similarity transformation (ST) for particle hole decoupling, IP/EA-UT-G$_0$W$_0$: Singles-doubles unitary transformation (UT) for particle hole decoupling, the degeneracy of the quasiparticle energies is shown in parentheses, def2-TZVP).}
		\label{tab:Decoupling2}
		\begin{tabular}{l | c c c | c c c} 
\hline
& \multicolumn{3}{c|}{F$_2$} & \multicolumn{3}{c}{MgO} \\
\hline
&  $\Delta$IP/EA-G$_0$W$_0$ & $\Delta$IP/EA-ST-G$_0$W$_0$ & $\Delta$IP/EA-$\Delta$UT-G$_0$W$_0$ & $\Delta$IP/EA-G$_0$W$_0$ & $\Delta$IP/EA-ST-G$_0$W$_0$ & $\Delta$IP/EA-UT-G$_0$W$_0$ \\
\hline
HOMO-2 & 2.978$^*$   &  $-$0.025     & $-$0.003     & 2.309      & 1.174       & 1.049\\
HOMO-1 & 1.833$^*$(2x) &  0.115(2x) & 0.166(2x) & 2.144      & 0.105       & 0.072\\ 
HOMO   & 2.083(2x) &  0.094(2x) & 0.134(2x) & 1.845(2x) &  0.341(2x) & 0.222(2x) \\
LUMO   & $-$2.120 & $-$0.091 & $-$0.067 & $-$0.801      & $-$0.006      & $-$0.029 \\
LUMO+1 & $-$0.355	 & $-$0.037 & $-$0.040 & $-$0.151(2x) & $-$0.028(2x) & $-$0.001(2x)\\
LUMO+2 & $-$0.367 & $-$0.058 & $-$0.056 & $-$0.130      & $-$0.022      & $-$0.023\\
\hline
    \hline
	\end{tabular}
 \\
\scriptsize{$^*$ Different energetic quasiparticle order for IP/EA-G$_0$W$_0$ compared to G$_0$W$_0$. The quasiparticle notation corresponds to the energetic order of G$_0$W$_0$.}
\end{table*}

\section{Conclusion and Outlook}

In this work we established a variety of exact relationships between the $\mathrm{G}_0\mathrm{W}_0$ approximation and  equation-of-motion (EOM) coupled-cluster theory within the quasi-boson formalism. The starting point was the exact equivalence between the $\mathrm{G}_0\mathrm{W}_0$ propagator and that of a particular electron-boson Hamiltonian supermatrix. From there, we could demonstrate the equivalence to a quasi-boson version of EOM-unitary coupled-cluster theory, and elucidate the differences, both theoretically, and numerically, from the standard EOM-similarity transformed coupled-cluster theory, as well as the recently introduced G$_0$W$_0$-TDA approximation. These relationships also motivated a particle-hole decoupling transformation of G$_0$W$_0$, which we formulated and numerically explored.  
\\
The precise relationships established here between the G$_0$W$_0$ approximation and various quantum chemistry theories, including CC theory, will be useful in future directions. For example, we did not focus on  computational costs in this work. However, the time-independent formulation of G$_0$W$_0$ opens up the incorporation of a wide variety of computational quantum chemistry techniques. It also suggests theoretical extensions to areas not traditionally treated using the GW formalism, such as multireference problems. The identification of unitary CC as the correct starting point to include all RPA screening in the quasiparticle energies also suggests new avenues to improve purely electronic theories of quasiparticle energies. 
\section*{Acknowledgments}
We thank T.~Berkelbach and S.~Bintrim for generously sharing their G$_0$W$_0$-TDA implementation.
We thank T.~Berkelbach and A.~Sokolov for helpful comments on the manuscript.
This work was supported by US Department of Energy through Award no. DE-SC0019390.
JT acknowledges funding through a postdoctoral research fellowship from the Deutsche Forschungsgemeinschaft (DFG, German Research Foundation) – 495279997. 
\section{Appendix}
\label{sec:appendix}
In this appendix some of the working equations of this paper are given in spin-free (sf) form.
For this, a singlet spin adaptation is used for the bosonic part \cite{toulouse2011closed}
\begin{align}
    b^\textrm{sf}_\mu = \frac{1}{\sqrt{2}}\left({b}^\alpha_\mu + b^\beta_\mu\right)
\end{align}
and fermionic creation and annihilation operators are replaced by
\begin{align}
    E^p_q = a^{\alpha,\dagger}_p a^\alpha_q + a^{\beta,\dagger}_p a^\beta_q.
\end{align}
The spin-free singlet $\mathbf{A}$ [Eq.~(\ref{eq:DefAMatrix})] and $\mathbf{B}$ matrix [Eq.~(\ref{eq:DefBMatrix})] are obtained as 
\begin{align}
    B_{\nu,\mu} &= B_{ia,jb} = 2 (ia|bj) \label{eq:BSpinFree}\\
    A_{\nu,\mu} &= A_{ia,jb} = (\epsilon_a - \epsilon_i)\delta_{ij}\delta_{ab} + 2 (ia|jb), \label{eq:ASpinFree}
\end{align}
$V_{pq\mu}$ as
\begin{align}
    V_{pq,\mu} = V_{pq,ia} = \sqrt{2} (pq|ia),
\end{align}
and $W_{pq\mu}$ as
\begin{align}
    W_{pq,\mu} = \sum_{ia} V_{np,ia} \left(\mathbf{X} + \mathbf{Y}\right)^s_{ia,\nu},
\end{align}
where $s$ refers to singlet excitations.
\subsection{Working equations IP+EA-EOM-CC}
\label{sec:ST-RPA}
The direct ring-CCD amplitudes are denoted as $t_{\nu \mu}^\textrm{rCCD}$. Denoting the quasiparticle amplitudes by $R$, then for IP+EA-qb-EOM-CCD, one finds the following $\sigma$ vector expressions
\begin{align}
    \sigma_i = \sum_{a\nu \mu}R_{a\nu}V_{ia\mu}t_{\nu \mu}^\textrm{rCCD},
\end{align}
\begin{align}
    \sigma_{\nu i} =&   \sum_{a\mu}R_{a}V_{\mu ia}t_{\nu \mu}^\textrm{rCCD}
 + \sum_{j\mu} V_{\mu ji}t_{\nu \mu}^\textrm{rCCD}R_{j} \notag \\
 &-\frac{1}{2} \sum_{\mu \lambda}B_{\mu \lambda}t_{ \nu \mu}^\textrm{rCCD}R_{\lambda i} - \frac{1}{2}\sum_{\mu \lambda}B_{\mu \lambda}t_{\nu \lambda}^\textrm{rCCD}R_{\mu i},
\end{align}
\begin{align}
    \sigma_a = \sum_{i \nu \mu}V_{\nu ba}t^\textrm{rCCD}_{\mu \nu} R_{\mu b},
\end{align}
\begin{align}
    \sigma_{\nu a} =&  \frac{1}{2}\sum_{\mu \lambda}B_{\lambda \nu}t^\textrm{rCCD}_{\mu \lambda}R_{\mu a} 
                    +  \frac{1}{2}\sum_{\mu \lambda}B_{\nu \lambda}t^\textrm{rCCD}_{\mu \lambda}R_{\mu a}.
\end{align}
\subsection{Similarity-transformed amplitude equations for IP/EA decoupling}
\label{sec:amplitudequations}
The singles amplitudes are obtained from
\begin{align}
 &f_{ia}
 - \sum_{j}f_{ji}t_{ja}
 + \sum_{b}f_{ab}t_{ib} 
 - \sum_{\nu j}W_{\nu ji}t_{\nu ja}\nonumber \\
 &+ \sum_{\nu b}W_{\nu ab}t_{\nu ib}
 - \sum_{jb}f_{jb}t_{ib}t_{ja} \nonumber \\ 
 &+ 2 \sum_{\nu jb}W_{\nu jb}t_{\nu ia}t_{jb}
 - \sum_{\nu jb}W_{\nu jb}t_{\nu ja}t_{ib} \nonumber \\ 
&- \sum_{\nu jb}W_{\nu jb}t_{\nu ib}t_{ja}
 = 0,
\end{align}
and the doubles amplitudes from
\begin{align}
&W_{\nu ia}
 + \sum_{\mu }\Omega_{\nu \mu}t_{\mu ia}
 - \sum_{j}f_{ji}t_{\nu ja}
 + \sum_{b}f_{ab}t_{\nu ib} \nonumber \\
 &- \sum_{j}W_{\nu ji}t_{ja}
  + \sum_{b}W_{\nu ab}t_{ib}
 - \sum_{jb}f_{jb}t_{\nu ja}t_{ib} \nonumber \\
  &- \sum_{jb}f_{jb}t_{\nu ib}t_{ja}
- \sum_{jb}W_{\nu jb}t_{ib}t_{ja}
  - \sum_{\mu jb}W_{\mu jb}t_{\nu ib}t_{\mu ja} \nonumber \\
 &+ 2\sum_{\mu jb}W_{\mu jb}t_{\mu ia}t_{\nu jb}
 - \sum_{\mu jb}W_{\mu jb}t_{\mu ib}t_{\nu ja} = 0.
\end{align}

\subsection{Commutator expressions $\bar{H}^\mathrm{eB}_\mathrm{uCC}$:}
\label{sec:CT-RPA}
The commutator expressions required for the construction of $\bar{H}^\mathrm{eB}_\mathrm{uCC}$ are
\begin{align}
    [\hat{V}^\mathrm{eB},\hat{\sigma}_\mathrm{B}] = \sum_{pq  \nu} \sum_\lambda V_{pq\lambda}  t_{\lambda \nu} \hat{a}^\dagger_p \hat{a}_q (\hat{b}^\dagger_\nu + \hat{b}_\nu)
\end{align}
and 
\begin{align}
    [\hat{H}^\mathrm{B},\hat{\sigma}_\mathrm{B}] = H^\mathrm{B}_0 + \tilde{H}^\mathrm{B}
\end{align}
with 
\begin{align}
    H^\mathrm{B}_0 = \mathrm{tr} \{\mathbf{Bt}\}
\end{align}
\begin{align}
    \tilde{H}^\mathrm{B} &= \frac{1}{2} \sum_{\nu \mu} \sum_\lambda  \left( A_{\nu\lambda}t_{\lambda \mu}  + t_{\nu\lambda}A_{\lambda \mu} \right) \hat{b}^\dagger_{\nu} \hat{b}^\dagger_{\mu} \notag \\ 
    &+  \frac{1}{2} \sum_{\nu \mu} \sum_\lambda  \left( A_{\nu\lambda}t_{\lambda \mu}  + t_{\nu\lambda}A_{\lambda \mu} \right) \hat{b}_\nu \hat{b}_\mu \notag \\
    &+ \sum_{\nu \mu} \sum_\lambda \left( B_{\nu\lambda}t_{\lambda \mu}  + t_{\nu\lambda}B_{\lambda \mu} \right) \hat{b}^\dagger_{\nu} \hat{b}_{\mu}.
\end{align}
\subsection{Abbreviations}
\label{sec:abbreviations}
A list of abbreviations of the different equation-of-motion (EOM) approaches considered in this work is provided in Tab.~\ref{tab:Abreviations}.
\begin{table*}
        \scriptsize
		\centering
		\caption{List of abbreviations of different equation-of-motion (EOM) approaches for the determination of charged excitations (ionization potential: IP, electron-affinity: EA) considered in this work (coupled cluster: CC).}
		\begin{tabular}{l  c c} 
    \hline
    Abbreviation & description & equation \\
    \hline
    G$_0$W$_0$ & G$_0$W$_0$ supermatrix & \ref{eq:DetailedMatrixform} \\
    G$_0$W$_0$-TDA & G$_0$W$_0$ supermatrix within the the Tamm-Dancoff approximation & \ref{eq:DetailedMatrixform1}\\
    IP+EA-EOM-CCSD & Similarity-transformed CC EOM using combined IP/EA operator basis & \ref{eq:IPEAsupermatrix}\\
    IP/EA-EOM-CCSD & Similarity-transformed CC EOM using separate  IP/EA operator basis & \ref{eq:eablocks},\ref{eq:ipblocks}\\
    IP+EA-qb-EOM-uCCD & IP+EA-EOM-uCCD in the quasi-boson approximation  & \ref{eq:UTBCC}\\
    IP+EA-qb-EOM-CCD & IP+EA-EOM-CCD in the quasi-boson approximation  & \ref{eq:STBCC} \\
    IP/EA-G$_0$W$_0$ & G$_0$W$_0$ without particle-hole coupling & \ref{eq:DetailedMatrixformReordered}\\
    IP/EA-ST-G$_0$W$_0$ & Singles-doubles similarity
transformation for particle-hole decoupling for G$_0$W$_0$ & \ref{eq:STDec}\\
    IP/EA-UT-G$_0$W$_0$ & Singles-doubles unitary
transformation for particle-hole decoupling for G$_0$W$_0$ & \ref{eq:UTDec}\\    
    \hline
	\end{tabular}
  \label{tab:Abreviations}
\end{table*}

\end{document}